\documentclass[apj]{emulateapj}
\slugcomment{{\sc Accepted to ApJ:} May 13, 2013} 
\usepackage{graphicx,amsmath,natbib}
\usepackage[flushleft]{threeparttable}

\begin{document}

\shorttitle{\uppercase{Solar Null-Point Particle Acceleration}}
\title{\uppercase{Kinetic modeling of particle acceleration in a solar null-point reconnection region}}

\shortauthors{G. Baumann, T. Haugb{\o}lle, \AA. Nordlund}
\author{G. Baumann$^1$, T. Haugb{\o}lle$^{2,1}$, and \AA. Nordlund$^{1,2}$}
\affil{$^1$Niels Bohr Institute, University of Copenhagen, Juliane Maries Vej 30, DK-2100 Copenhagen, Denmark \\
$^2$Centre for Star and Planet Formation, Natural History Museum of Denmark, University of Copenhagen,
{\O}ster Voldgade 5-7, DK-1350 Copenhagen, Denmark
}
\email{gbaumann@nbi.ku.dk}

\newcommand{\fig}[1]{Fig.\ \ref{fig:#1}} 
\newcommand{\Fig}[1]{Figure \ref{fig:#1}} 
\newcommand{\bea}[1]{\begin{eqnarray}\label{eq:#1}}
\newcommand{\eea}{\end{eqnarray}}
\newcommand{\beq}[1]{\begin{equation}\label{eq:#1}}
\newcommand{\eeq}{\end{equation}}
\newcommand{\Eq}[1]{Equation~\ref{eq:#1}}
\newcommand{\Eqs}[2]{Equations~\ref{eq:#1}--\ref{eq:#2}}
\newcommand{\Section}[1]{Section~\ref{sec:#1}}
\newcommand{\Table}[1]{Table~\ref{table:#1}}
\newcommand{\ddt}[1]{\frac{\partial #1}{\partial t}}
\newcommand{\ddx}[1]{\frac{\partial #1}{\partial x}}
\newcommand{\ddy}[1]{\frac{\partial #1}{\partial y}}
\newcommand{\ddz}[1]{\frac{\partial #1}{\partial z}}
\newcommand{\ddi}[1]{\frac{\partial #1}{\partial r_i}}
\newcommand{\ddj}[1]{\frac{\partial #1}{\partial r_j}}
\newcommand{\DDt}[1]{\frac{D #1}{ D t}}

\providecommand{\e}[1]{\ensuremath{\times 10^{#1}}}
\renewcommand{\div}{\nabla\cdot}
\newcommand{\curl}{\nabla\times}
\newcommand{\grad}[1]{{\nabla #1}}
\newcommand{\uu}{\mathbf{u}}
\renewcommand{\gg}{\mathbf{g}}
\newcommand{\pp}{\mathbf{p}}
\newcommand{\JJ}{\mathbf{J}}
\newcommand{\BB}{\mathbf{B}}
\newcommand{\EE}{\mathbf{E}}
\newcommand{\ux}{u_{\rm x}}
\newcommand{\uy}{u_{\rm y}}
\newcommand{\uz}{u_{\rm z}}
\newcommand{\gz}{g_{\rm z}}
\newcommand{\Hp}{H_{\rm P}}
\newcommand{\rr}{\mathbf{r}}
\newcommand{\Om}{\mathbf{\Omega}}
\newcommand{\Frad}{\mathbf{F}_{\rm rad}}
\newcommand{\Qrad}{Q_{\rm rad}}
\newcommand{\Qvisc}{Q_{\rm visc}}
\newcommand{\Qjoule}{Q_{\rm Joule}}
\newcommand{\Tvisc}{\mathbf{\tau}_{\rm visc}}
\newcommand{\Tij}{\tau_{ij}}
\newcommand{\sij}{s_{ij}}
\newcommand{\Bv}{B_{\nu}}
\newcommand{\Sv}{S_{\nu}}
\newcommand{\Iv}{I_{\nu}}
\newcommand{\tauv}{\tau_{\nu}}
\newcommand{\kv}{\kappa_{\nu}}
\newcommand{\Ekin}{E_{\rm kin}}
\newcommand{\Etherm}{E}
\newcommand{\Fconv}{\mathbf{F}_{\rm conv}}
\newcommand{\Fvisc}{\mathbf{F}_{\rm visc}}
\newcommand{\Fkin}{\mathbf{F}_{\rm kin}}
\newcommand{\half}{\frac{1}{2}}
\newcommand{\lbc}{{\rm bot}}

\begin{abstract}
The primary focus of this paper is on the particle acceleration mechanism in
solar coronal three-dimensional reconnection null-point regions.
Starting from a potential field extrapolation of a Solar and Heliospheric Observatory (\textit{SOHO}) magnetogram taken on
2002 November 16, we first performed magnetohydrodynamics (MHD) simulations with horizontal motions
observed by \textit{SOHO} applied to the photospheric boundary
of the computational box. After a
build-up of electric current in the fan-plane of the
null-point, a sub-section of the evolved MHD data was used as initial and
boundary conditions for a kinetic particle-in-cell model of the plasma.
We find that sub-relativistic electron
acceleration is mainly driven by a systematic electric field in the current sheet. A
non-thermal population of electrons with a power-law distribution in energy
forms in the simulated pre-flare phase, featuring a power-law index of about -1.78.
This work provides a first step towards bridging the gap between macroscopic scales
on the order of hundreds of Mm and kinetic scales on the order of cm in
the solar corona, and explains how to achieve such a cross-scale coupling by utilizing
either physical modifications or (equivalent) modifications of the constants of
nature. With their exceptionally high resolution --- up to 135 billion
particles and 3.5 billion grid cells of size 17.5\,km --- these simulations
offer a new opportunity to study particle acceleration in solar-like settings.
\end{abstract}

\subjectheadings{acceleration of particles – magnetic reconnection – Sun:  fares – Sun: corona}

\section{Introduction}
\label{sec:introduction}

During solar flares an enormous amount of energy is released, in particular in
the form of highly energetic non-thermal electrons. It is generally accepted that
the underlying release mechanism is magnetic reconnection. In connection with a
reconnecting current sheet, strong large-scale electric fields can build up
(\cite{2009ApJ...690.1633L} and references therein), leading to direct
acceleration of particles beyond thermal velocities, while fluctuating electric
fields created by reconnection and other dynamic events can lead to stochastic
acceleration
\citep[e.g.][]{1996ApJ...461..445M,1997JGR...10214631M,2006ApJ...644..603P}.

During the last decade, high-resolution observations of solar magnetic fields
from several solar space missions and ground-based observations, such as \textit{YOHKOH}, \textit{SOHO},
\textit{TRACE}, \textit{RHESSI}, \textit{HINODE}, \textit{STEREO}, \textit{Solar Dynamics Observatory} and \textit{Spitzer Space Telescope}
\citep[e.g.][]{2005ApJ...630..596L, 2003ApJ...594.1068K, 2006ApJ...638L.117M, 1999SoPh..190....1P,
2004ApJ...612..546S, 2007Sci...318.1588A, 2012ApJ...746...67K, 2008ApJ...688L.119J}
have brought new insights regarding the interconnecting and X-ray loops, the current sheet in the reconnection region, upflow velocities of
chromospheric evaporation as a result of non-thermal particles impact, and
white-light emission during flares and instabilities as possible drivers,
providing substantial support for existing solar flare and coronal mass ejection models.


Additionally, new three-dimensional (3D) coronal magnetic reconnection and
acceleration models \citep[e.g.][]{1996RSPSA.354.2951P, 2004ApJ...608..540V} as
well as the rapid expansion of computing resources for large-scale simulations
\citep[]{2007ASPC..369..355I,2011A&A...529A..20G,2012A&A...539A..22T} have opened up a new chapter
in the understanding of the formation of current sheets and particle
acceleration sites in 3D reconnection regions. But what is
lacking is an interconnected understanding of how microscopic plasma physics scales
interact and exchange information with macroscopic large-scale
magnetohydrodynamics (MHD) scales in the solar atmosphere. MHD scales provide
the environment for particle acceleration to happen, but it is unclear how the
complex non-linear feedback from much smaller scales onto the overall behavior
of the plasma above the solar surface is handled by nature.

Most studies have made use of the fact that the temporal evolution of the
large-scale magnetic field in the solar atmosphere can to a first approximation
be described by compressible MHD. But fluid approaches are limited to thermal
particle distributions, which is not sufficient to describe the kinetic aspects
of magnetic reconnection that convert magnetic field energy into particle kinetic
energy. A proper description of such processes
requires taking into account the back-reaction of kinetic
processes on the large-scale dynamics. Nevertheless, test particle MHD
simulations,
\citep[e.g.][]{2005ApJ...620L..59T,2006A&A...449..749T,2010A&A...520A.105B,2005A&A...436.1103D,2008A&A...491..289D,2010A&A...511A..73R},
give a good idea of the overall acceleration region framework. In such
simulations, MHD fields evolve independently of the motion of the test
particles. There is no immediate back-reaction from the accelerated particles onto the
fields, which is potentially a serious limitation, since especially the changes of the
electric field that would be
induced by the accelerated particles can be large compared to the background
field induced by magnetic reconnection \citep{2009JPlPh..75..619S}. The lack of feedback can
lead to an exaggerated particle acceleration, as has been noted by, for example,
\citet{2010A&A...511A..73R}, since there is no limitation for the energy gain
of particles. Furthermore, kinetic instabilities can be of importance for the
fast reconnection onset in solar flares, and more generally for the evolution of
the current sheets in reconnection regions \citep{2010AdSpR..45...10B}.
There is therefore a need for realistic self-consistent kinetic simulations to
examine micro-scale processes in plasmas and to be able to properly take into
account back-reactions from the particles to the fields.

The main challenge for kinetic simulations on scales of solar events
is the enormous dynamic range involved. Explicit particle-in-cell (PIC)
simulations have to resolve characteristic kinetic scales and are restricted to
very small
physical sizes, due to limitations that arise from conditions for code stability
as well as from resolution criteria.
These computational restrictions have so far prevented investigations of the coupling of kinetic
to MHD scales using kinetic simulations, and have thus prevented self-consistent
modeling of particle acceleration in solar flares. There has mainly been one
interlocked model attempt by \citet{2007JCoPh.227.1340S}, who ran PIC simulations embedded in a large-scale MHD simulation with PIC boundary conditions defined by the surrounding MHD domain. The particle information was similarly passed from the MHD to the kinetic simulation areas as it is done in the present study. But the
complexity of the problem especially in the transition zone between MHD and PIC zones combined with the simultaneity of the two simulations limits the
applicability of such attempts significantly. 
In the present study, we introduce a new method to meet the challenge of
multi-hierarchy simulations and use the method to investigate particle
acceleration mechanisms in a solar reconnection pre-flare event using ultra-large-scale kinetic modeling. We further discuss the limitations of this new way to combine macroscopic and microscopic scales.

In \Section{methods}, we describe the numerical methods and their
implementation, while in \Section{simulations} we introduce the experimental
setup and the necessary modifications and list the most important simulations we performed, indicating their
relative roles and importance. In \Section{results}, we present and discuss the
results including a comparison to observational studies. Finally, in \Section{conclusions}, we summarize the results, present our
conclusions, and give an outlook onto future work.

\section{Methods}
\label{sec:methods}
We perform PIC simulations using the \emph{Photon-Plasma}
code  \citep{Haugboelle:2005,Hededal:2005b}, which solves the Maxwell equations
\bea{maxwell}
-\ddt{\BB} &=& \curl \EE \\
\epsilon_0 \mu_0\ddt{\EE} &=& \curl \BB - \mu_0\JJ ,
\eea
together with the relativistic equation of motion for charged particles
\bea{equ:Lorentz}
m\frac{d(\gamma \bf{v})}{dt} &=& q(\EE + \bf{v} \times \BB)
\eea
on a staggered Yee lattice \citep{1966ITAP...14..302Y}. We use SI units,
scaled so the unit of length is 1\,km, the unit of time is 0.1\,s, and the unit
of density is $10^{-12}$ kg m$^{-3}$.

The Lorentz force is computed by interpolation of the electromagnetic fields $\EE$ and $\BB$
from the mesh to the particle positions, employing a cubic scheme using the 64
nearest mesh points. The code integrates the trajectories of protons and electrons
moving in the electromagnetic field with a Vay particle mover \citep{2008PhPl...15e6701V}.
The charge density $\rho$ is determined by weighted averaging of the particles
to the mesh points using the same cubic scheme as for the field to particle
interpolation, to avoid self forces. The currents are found
using a new 6$^\textrm{th}$ order version of the Esirkepov charge
conservation method \citep{2001CoPhC.135..144E} that is consistent with the field
solver\footnote{The original Esirkepov method, just like most other charge conserving
schemes in the literature, is only consistent with a 2$^\textrm{nd}$ order field solver.}.
The field equations are solved on the mesh using an implicit  2$^\textrm{nd}$ order
in time and 6$^\textrm{th}$ order in space method. Because the solver is charge
conserving and the fields are properly staggered, Gauss' law is obeyed and the $\BB$ field is
kept divergence free to numerical precision.

The boundaries of the domain are fixed for the magnetic fields and open for
particles. Particles can escape and new particles are added to the box from
`ghost cells' outside the physical boundaries, where the conditions are
specified from values in the MHD snapshot.  In the current short-duration
kinetic simulations, the field values in the boundaries are held fixed.  In longer duration
simulations, they could be made time dependent by performing
interpolations in time between MHD snapshots. See also \citet{2012arXiv1211.4575H}.

Below we introduce for the first time a modification of the elementary charge $q$,
in addition to and analogous to the well-known speed of light ($c$) modification that has previously
been used in many cases \citep[e.g.][]{2006Natur.443..553D}. Since all the micro-scales
(gyro-radii, skin depths, Debye lengths) are inversely proportional to the charge
per particle, one can increase the micro-scales until they are resolvable on macroscopic
scales by decreasing the charge per particle sufficiently.

While changing the ratio of micro- to macro-scales by a large amount may appear
to be a very drastic approach, the method can be defended on both qualitative
and quantitative grounds:  from a qualitative point of view, it may be argued
that as long as one retains a proper ordering of non-dimensional parameters, as
discussed in more detail below, one should expect to see essentially the same
qualitative behavior, albeit with (possibly large) differences in quantitative
aspects. From a quantitative point of view, one can indeed attempt to predict how
these quantitative aspects depend on the modifications of $q$ and $c$, to be able
to extrapolate---at least to order of magnitude---to values that would be
typical in the unmodified system.

Adopting this approach enables us to perform explicit PIC
simulations of large-scale plasmas. With the parameter values used here we
resolve the electron skin depth $\delta_{skin,e}$ with at least 3.8 grid cells and
the Debye length with at least 0.3 grid cells.
For the time stepping, we use a Courant condition of 0.4,
considering the light crossing time in a grid cell as well as the local plasma
frequency.

\section{Simulations}
\label{sec:simulations}
In order to investigate the particle acceleration mechanism around a 3D
reconnection region, we started out with a Fourier transform potential
extrapolation of a \textit{SOHO}/Michelson Doppler Imager
magnetogram from 2002 November 16 at 06:27:00\,UT, 8 hour prior to a
C-flare occurrence in the AR10191 active region. This is similar to the setup by
\citet{2009ApJ...700..559M}. As there is no vector
magnetogram available for this event, a non-linear force-free extrapolation has
not been feasible.
The potential field magnetic configuration arising from the extrapolation
showed mainly two connectivity regions, separated by a dome-shaped `fan-surface'
\citep{1997ApJ...485..383C}, each including a spine structure, which is the
symmetry line intersecting the fan at the magnetic null-point. Although initially strove for, the given setup did not lead to a flare-like plasma eruption. The MHD simulation hence describes a pre-flare phase with a characteristic 3D magnetic reconnection region.

\begin{figure}
\begin{center}
    \includegraphics[width=1.0\linewidth]{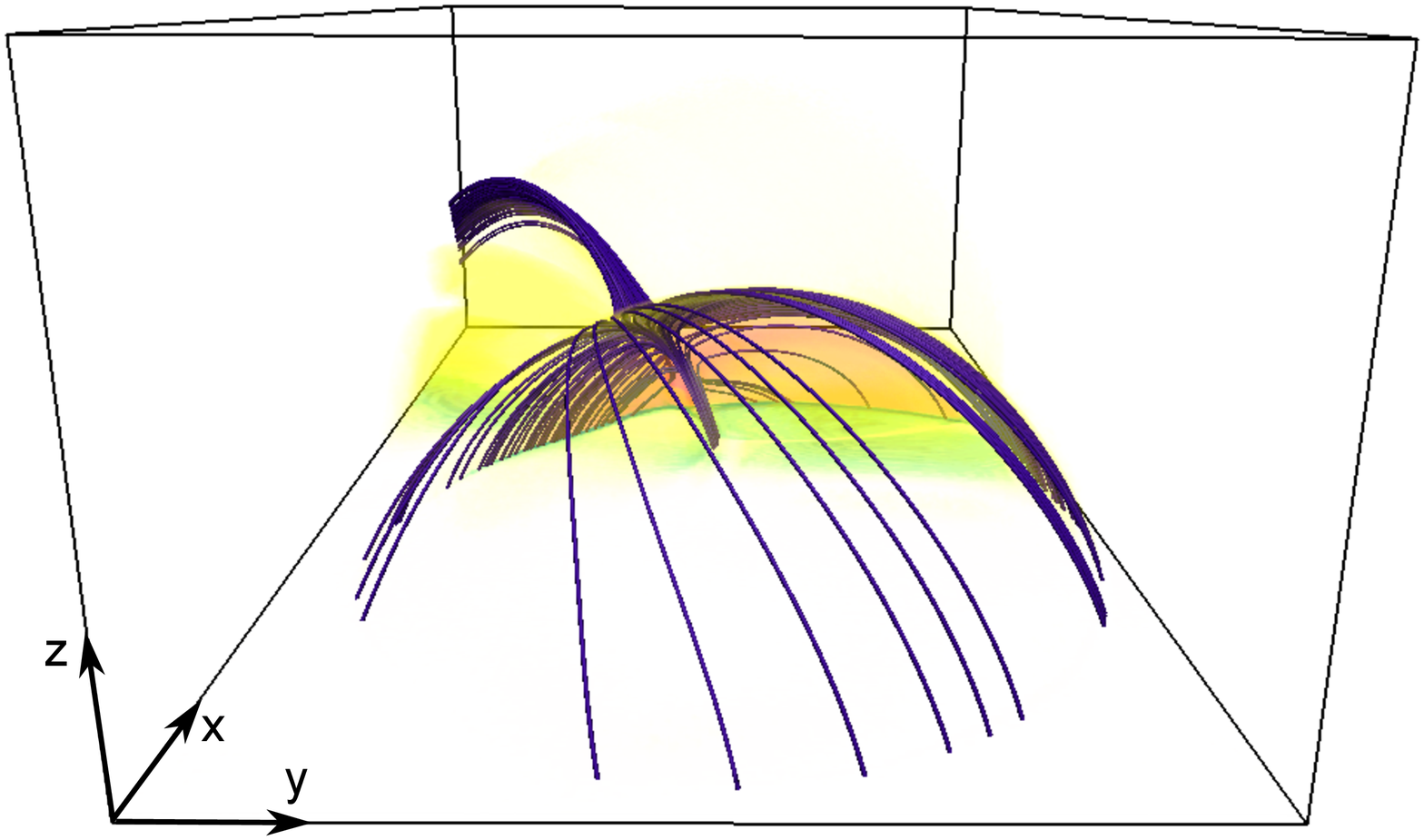}
(a)
    \includegraphics[width=1.0\linewidth]{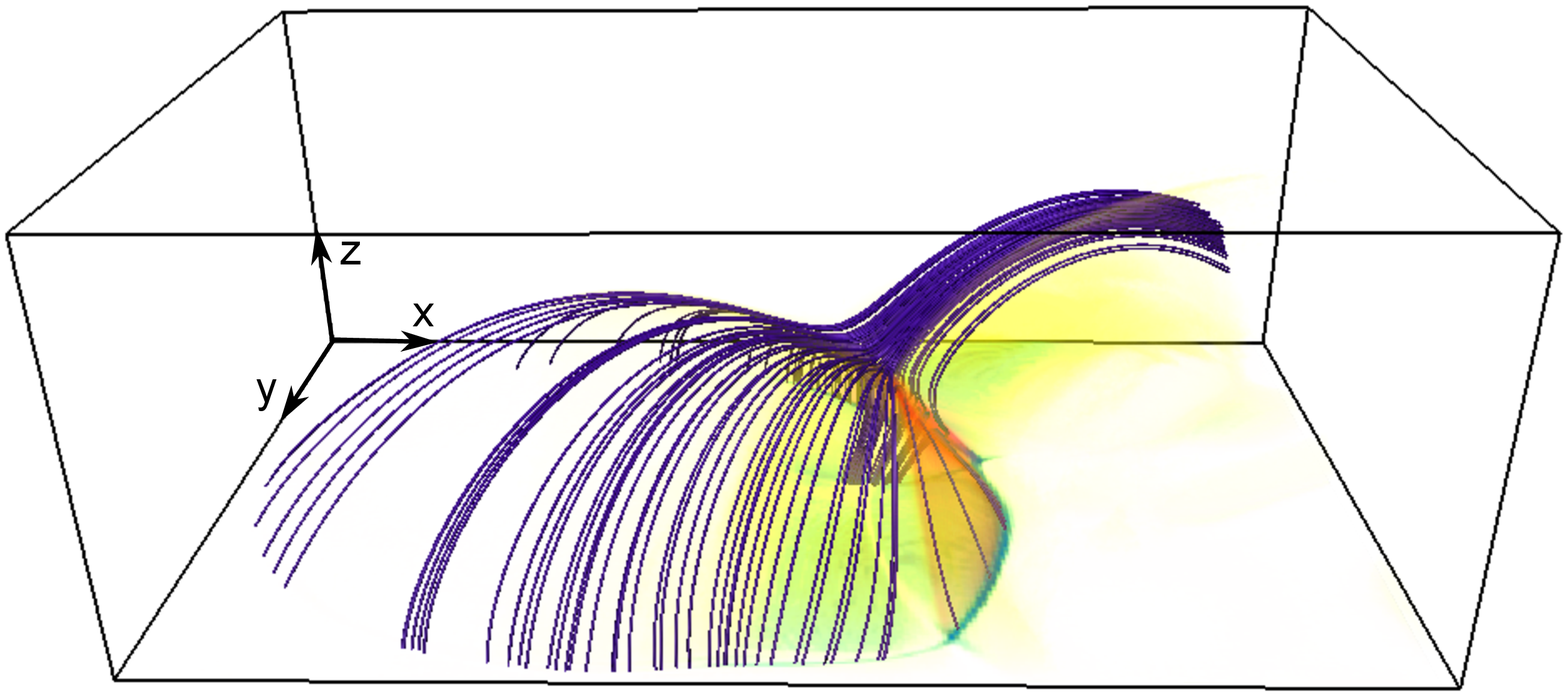}
(b)
\end{center}
  \caption{Initial state of the PIC code simulation; a cut-out of size
44$\times$25$\times$16\,Mm of an evolved MHD simulation snapshot. (a) Seen
along the \textit{x}-axis and (b) along the \textit{y}-axis. Shown are magnetic field lines
(purple), the maximum current density at the lower boundary (turquoise), and the
electric current
density as a semitransparent direct volume rendering (yellow for lowest value, and
orange for highest value).}
    \label{fig:currentDVRandPROBE_bfieldFLOW}
\end{figure}

We define the
inner spine as the spine intersecting with the solar surface inside the
dome-like surface delimited by the fan-plane, while the outer
spine reaches up into the corona before returning to the photosphere several
tens of Mm away from the null-point (cf. Figure \ref{fig:currentDVRandPROBE_bfieldFLOW}).
The magnetic field lines in the fan-plane meet at the magnetic null-point,
which is located at about 4\,Mm above the photospheric boundary.\footnote{For
computational reasons, we tapered off the vertical magnetic field toward the
boundaries of our horizontally periodic MHD model, and the height of the
null-point is therefore different from the case in
\citet{2009ApJ...700..559M}.} The dome-like fan-plane is a typical feature of
 a parasitic polarity magnetic null-point topology
originating when a vertical dipole field emerges into a magnetic field
configuration of opposite magnetic polarity. For an elaborated description of
the observations of this solar flare event, see \citet{2009ApJ...700..559M}.

The MHD simulations of this event previously conducted using the fully 3D resistive compressible \emph{Stagger} MHD code \citep{Nordlund1997} on a box size of
$175\times100\times62$\,Mm and a stretched $896\times512\times320$ grid with a
minimum cell size of $\approx$ 80\,km in a region around the null-point
\citep{2013SoPh..284..467B} establishes the initial condition for the PIC simulations.
In most cases, the initial MHD particle density and temperature were set to
the constant values of
$6.8\e{12}$\,cm$^{-3}$ and $5\e{5}$\,K, neglecting gravity, analogously to
\citet{2009ApJ...700..559M}. The boundaries were
chosen to be closed at the bottom
and top, and periodic on the sides. Assuming, as did \citet{2009ApJ...700..559M},
that the strong horizontal motions observed
by \textit{SOHO} during the considered time interval was the driver of the reconnection
event, we simulated the plasma motion caused by these motions by using essentially
the same elliptically shaped driver at
the photospheric boundary of the box as was used by
\citet{2009ApJ...700..559M}, applying velocity amplitudes ranging from
3.33 to 20\,km s$^{-1}$ and a horizontal shape function schematically
representing the observed photospheric motion. The enhanced driving speed,
compared to observations, led to a significant reduction of the simulation
time, while it was still sub-Alfv\'enic and thus left the overlying plasma and
magnetic field enough time to adapt to the displacement.

The line-tied photospheric boundary motions indirectly reshuffled the fan-spine geometry at
its foot-points by a pressure increase onto the fan-plane, which caused a
relative displacement of the magnetic field lines inside and outside the
fan-plane of the magnetic null-point, respectively, resulting in the
formation of a current sheet in the fan-plane.  The displacement of field lines
also causes a relative displacement of the two spines, which then leads
to magnetic reconnection at the null-point and a growing resistive electric
field in the fan-plane. 
A detailed description of the MHD simulations and their results are published
in a companion paper \citep{2013SoPh..284..467B}.

For the present study, we used an MHD simulation with an initial particle
density of $6.8\e{12}$\,cm$^{-3}$, a temperature of $5\e{5}$\,K, and an applied
driving speed of 20\,km\,s$^{-1}$. After 240 seconds of simulated solar time,
the null-point area showed clearly
enhanced current densities parallel to the magnetic field, indicating that a
dissipative process was taking place. We choose this as the starting point for
the 3D relativistic PIC simulation. To minimize computational constraints due
to the plasma frequency, we rescale the density from the essentially chromospheric
value of $6.8\e{12}$\,cm$^{-3}$ used by \citet{2009ApJ...700..559M} partway
toward values more characteristic of an active region corona, so that it
becomes instead $1.28\e{11}$\,cm$^{-3}$.

Figures \ref{fig:currentDVRandPROBE_bfieldFLOW}(a) and
(b) illustrate the initial PIC simulation
setup, by showing the chosen cut-out of a snapshot from the MHD simulation, as
seen along the \textit{x} and \textit{y} axis respectively. The outer spine extends to the right
of Figure \ref{fig:currentDVRandPROBE_bfieldFLOW}(b), while the fan-surface
spreads out over the rest of the area. Due to the initial difference in the
fan-plane eigenvalues \citep{2009ApJ...700..559M}, a slight
asymmetry in the fan-surface is noticeable.  The inner spine can be recognized
inside the volume spanned by the fan-plane field lines in Figure
\ref{fig:currentDVRandPROBE_bfieldFLOW}(a).

For the kinetic simulations, we use the \emph{Photon-Plasma} code on a
uniform grid. A cut-out of the MHD simulation box with dimensions
44$\times$25$\times$16\,Mm is chosen for a set of simulations, covering
cell sizes from 70\,km down to 17.5\,km corresponding to uniform grids with up to
2518$\times$1438$\times$923 cells. The simulations are generally performed with
20 particles per species per cell, covering up to 30 solar seconds. We neglect
gravitation, since the aim of this study is mainly to assess the electron
acceleration mechanism, for which the influence of the gravitational force is
negligible. The initial ion velocities consist of two components; a random
thermal velocity drawn from a Maxwellian distribution plus the bulk velocity
from the MHD simulation. For the electron velocities, we use the sum of a
random thermal velocity drawn from a Maxwellian distribution, the bulk velocity,
and the velocity due to the initial electric current. The initial electric
field is simply taken as the advective electric field ($-\uu\times\BB$, where $\uu$ is the MHD bulk velocity) from the MHD simulation,
which will be self-corrected by the system after the first few time-steps of
the kinetic simulation, by adapting the electric field as well as the electric
current to balance the system.

\subsection{Modifications to Resolve Kinetic Scales}
It is common in MHD and PIC simulations to convert physical units into
numerically more convenient code units. Such a rescaling
leaves the simulated physical situation completely unaffected and all parameters
can easily be rescaled back to physical units at any time.
Because they leave the ratio of scales unaffected these kinds of
conversions between units of measurement are not sufficient to allow
explicit PIC codes to address the multi-scale issues discussed below.

Magnetic reconnection is a typical example of multi-scale physics, where
microscopic scales interact with and couple to macroscopic scales. While
kinetic scales in the solar corona are on the order of mm, coronal structures
such as the ones investigated here have scales on the order of tens of Mm. This
range of spatial scales is impossible to cover by explicit kinetic simulations
--- now as well as in the foreseeable future.  Explicit PIC codes are subject to
a number of numerical stability constraints. For the code that we use here,
one of them is that the Debye
length, defined as $\lambda_D = \sqrt{\epsilon_0 k_B T / n_e q_e^2}$, where
$k_B$ is the Boltzmann constant, $T$ the temperature, $\epsilon_0$ the vacuum
permittivity, $n_e$ the electron density, and $q_e$ the electron charge,
 should be at least on the order of a fraction of a grid cell $\Delta s$:
\begin{equation}
    0.3\;\Delta s \lesssim \lambda_D .
    \label{equ:debye}
\end{equation}
In addition, there are also stability constraints on the time step set
by the speed of light and the plasma frequency
\begin{equation}
    \Delta t < \Delta s / c ,
    \label{equ:debye2}
\end{equation}
\begin{equation}
    \Delta t \lesssim 2 /\omega_p .
    \label{equ:debye3}
\end{equation}
In summary, code stability requires that the time step needs to approximately
resolve the light-wave and plasma-wave propagation and that the grid spacing
should not severely under-resolve the electron Debye length.  When these conditions
are marginally fulfilled in the global simulation domain, they are typically
(except for the speed of light condition) fulfilled with a good margin in most
computational cells, which helps to improve the accuracy of the numerical
solution.

Additionally, we need to fulfill the following approximate equality:
\begin{equation}
 \delta_{skin} \approx \mbox{a few} \cdot \Delta s.
 \label{equ:skin}
\end{equation}
This requirement arises to ensure a reasonably faithful representation of the
plasma, as employing too few grid cells per skin depth suppresses sub-skin depth
plasma behavior,
while, for a fixed number of grid cells, having too many grid cells per skin
depth results in a model where the kinetic scales and the MHD scale are not
sufficiently separated, and consequently, the MHD behavior is lost.
Given these constraints, we would require approximately one septillion
($1\e{24}$) grid cells for an unmodified simulation to simultaneously resolve the
microscopic and macroscopic scales in the problem. This fundamental constraint
notwithstanding, bridging the large-scale difference is in fact approachable. One can
either change some of the physical properties (e.g.~the magnetic field strength
or the plasma density) or, equivalently, modify some of the constants of nature.
As long as such
modifications of the physical properties of the system do not change the large-scale behavior, they may be helpful by decreasing the gap between
microscopic and macroscopic scales.

We first introduce two possible sets of modifications of the physical
properties which conduce to decrease the large gap between the micro-scales and the
macro-scales, using three modification parameters, which are closely related to
each other, so that only two of them are so to speak free parameters, while the
third one is a result thereof. One of the reasons for this is an MHD variable
constraint; we wish to maintain the relative importance of magnetic and fluid
pressure, i.e., we wish to keep the plasma beta parameter unchanged in each type of
modification. The goal is to change micro-scales, which cause the previously
mentioned numerical constraints, Equations\,(\ref{equ:debye})--(\ref{equ:skin}), i.e.~increasing the Debye length $\lambda_D$,
decreasing the span of velocities (equivalent to decreasing the speed of
light), and decreasing the plasma frequency. In modification of type A we change the
magnetic field strength by a factor $b < 1$, implying a change in the density to
keep the pressure balance. This increases the length scales
($\lambda_D$, the gyro-radius $r_g$ and the skin depth $\delta_{skin}$) and at
the same time reduces the frequencies ($\omega_p$ and the gyro-frequency
$\omega_g$) by the same factor, while leaving the temperature and hence the
sound speed $v_s$ and Alfv\'en speed $v_A$ of the system unchanged.

Modifications of type B address the macroscopic speed ratios. We start
by changing the temperature by a factor $t^2$, which then leads to an increase
in the acceleration of gravity $g$ by the same factor, as the scale height is proportional to $T$/$g$. The consequences of
this modification are that speeds increase by a factor $t$, while the length and frequency
ratios are forced to be changed to achieve these required modifications, which
means that for this type of change one cannot maintain the micro-scale ratios.

These modifications of physical properties are equivalent to changing the
constants of nature $q$ and $c$; the unit of charge and the speed of light.
Then $q^{-1}$ determines the
ratio between micro and macro-scales through the charge density and $c^{-1}$
influences the ratio of the speeds. $\mu_0$ needs to be kept constant, in order
to keep the magnetic pressure and hence the plasma beta unchanged, hence
$\epsilon_0$ is the parameter that needs to be modified.
Here is a summary of these two types of modifications:
\begin{enumerate}
\item[A:] \uppercase{Physical changes}: $B\sim b, \rho\sim b^2, T = \textrm{constant}$\\
\uppercase{Consequences}: $ \lambda_{D}\sim \delta_{skin}\sim r_{g}\sim b^{-1},
\omega_p\sim \omega_g\sim b, v_A\sim v_s\sim v_{th} = \textrm{constant}$
\item[B:] \uppercase{Physical changes}: $T\sim t^2, g\sim t^2, B\sim t, \rho = \textrm{constant}$ \\
\uppercase{Consequences}: $\lambda_{D}\sim t, \textrm{ but } \delta_{skin}\sim \omega_p
= \textrm{constant}, v_A\sim v_s\sim v_{th}\sim t$.
\end{enumerate}

The idea of modifying the constants of nature has previously been
employed. \citet{2006Natur.443..553D} reduced in their approach the speed of light
$c$, while \citet{2009JPlPh..75..619S} reduced the particle
density, which has the same effect on the micro-scales as lowering the charge
per particle. In fact, one can show that changes of on the one hand
temperature and particle density, and on the other hand changes of the
speed of light and the charge per particle, have similar effects:  decreasing
the speed of light or increasing the temperature (squared) both reduce the
ratio of the speed of light to the thermal speed, and decreasing the particle
density or the charge per particle both increase the ratio of micro- to
macro-scales (adjusting at the same time the magnetic field so as to keep
the plasma beta the same). So, in that sense, our changes of the speed of
light and the electric charge per particle effectively corresponds to simulating a
coronal region with a very high temperature and a very low particle density.

Apart from the modifications discussed above, a reduction of the ratio of the
electron to proton mass from 1836 to 18 is used. This is done to decrease the
gap between the ion and electron plasma frequency and skin depths to acceptable
values. It is a standard trick in PIC simulations,
and mass ratios above 16 are normally considered enough to separate the two scales.

Decreasing the ratio between the different speeds (e.g. by lowering the
speed of light) are generally motivated by an assumption that
if the speeds are nevertheless much smaller than the speed of light, one
expects only very marginal changes in the dynamics, while the savings in
computing time (which scales as the speed of light) are considerable.

Changing the charge per particle with a large amount (on the order of $10^6$
here) is more drastic, but is necessary to bring micro-scales into a realm
that is resolvable with current computing resources.  To lowest order, such
a change of scales is not expected to change the magnetically dominated
dynamics dramatically; charged particles are still forced to move essentially
along magnetic field lines, with gyration orbits oriented in essentially the
same manner with respect to the large-scale structures.  What changes is
exactly what is required to change, namely, the ratios between micro- and
macro-scales. Any effect that depends on this ratio then changes in an
in principle predictable fashion, and one can, a posteriori, attempt to compensate
for this, when analyzing and discussing the results, as is done below in Section \ref{sec:DCefield}.

A crucial point when modifying the constants of nature is to
ensure that the hierarchy of characteristic speeds, times, and length scales
are, to the largest extent possible, kept as in the real average coronal
environment.  For the characteristic speeds, this hierarchy is
\begin{equation}
v_d < v_{th,p} \sim v_{s} < v_{th,e} < v_A < c ,
\label{equ:scaling1}
\end{equation}
where $v_d$ is the average electron drift speed that arises due to the electric
current, $v_{th,p(e)}$ is the thermal speed of protons(electrons), $v_s$
is the speed of sound, $v_A$ is the Alfv\'en speed, and $c$ is the speed of
light.
For the corona, the values of these inequalities can be approximated by the
following numbers (km s$^{-1}$):
\begin{equation}
0.002 < 90 < 3\,900 < 12\,000 < 300\,000.
\label{equ:scaling2}
\end{equation}
In contrast to MHD simulations of a solar event, the magnitude of
the parameters of a scaled PIC simulation cannot be directly compared
to observationally obtained values, since the modifications of the
constants of nature have an influence on the magnitudes of the parameters.
But estimating what the corresponding solar values would be is possible,
by careful rescaling.

We measure all quantities in scaled SI units (unit of
length = 1 km, unit of time = 0.1 s, and unit of mass density = $10^{-12}$
kg\,m$^{-3}$).

\subsection{Modification Validation\label{sec:validation}}
To analyze the influence of the presented modifications onto the physical processes at work, we performed several simulation runs with different physical resolutions and modifications of $q$ and $c$.
A subset of the relevant simulations may be found in Table \ref{tab:SimulationRuns} which summarizes the typical initial run
parameters for each simulation.
 
Run 1S is the smallest run, which therefore corresponds to the size category \textit{S}, while category \textit{M} runs have double the physical resolution and \textit{L} simulations correspondingly a four times higher physical resolution. Runs 1S, 3M and 5L have a constant ratio of the cell size per kinetic scale and hence test the numerical convergence of the simulation results. Run 5L is of special interest, due to its drift speed that is lower than the thermal electron speed, as is the case in the real Sun. Furthermore, a comparison of runs 3M and 2M shows the effects of a change of the micro-scales while keeping the resolution constant. These two case comparisons are especially interesting when considering the kinetic energy distribution for high-energetic particles as, e.g. power-law indices can be validated against observations. In both cases, as presented in Section \ref{sec:distr}, the distributions are, despite their differences in the modification of the constants of nature, very similar and, as mentioned in Section \ref{sec:conclusions}, their power-law indices are comparable to observational results. This indicates, that the influence of the chosen numerical resolutions onto the physical acceleration process of particles in the reconnection region is small. Run 4M is specifically interesting when looking at the upper energy limit of the kinetic energy distribution of particles, as it shows the influence of an increase in the speed of light onto the energy gain of particles. Very little difference in the energy distribution could be found, which attests the reduction of the speed of light as a reasonable modification as long as one conforms to the speed hierarchy. We also increased the systems temperature by a factor of two for run 2M and run 3M resulting in run 2M$_T$ and run 3M$_T$. This increase in temperature causes the high-energy tail to (partially) drown in the thermal particle energy distribution, so that a study of the nature of non-thermal particle acceleration in the chosen magnetic field geometry is no longer possible as no clear distinction between the two energy ranges can be made.

The results of these modification studies are in more detail reported in Section \ref{sec:distr}. Section \ref{sec:DCefield} additionally discusses the resulting electric fields of most simulation runs, listed in Table \ref{tab:SimulationRuns}. In summary, we note that except for the under-resolved run 1S, the electric fields of simulations with different modifications cover a rather small range of values.

\begin{table*}
    \begin{threeparttable}
    \centering    
    \caption{Summary of Simulation Runs.\label{tab:SimulationRuns}}
        \begin{tabular}{l c c c c c c c c c c}
            \tableline\tableline
            \\
 Run & $\Delta$s (km) &$\delta_e/\Delta$s & $\Delta$t & $r_g/\Delta$s & $\lambda_D/\Delta$s & $c$ (km
s$^{-1}$) & v$_J$ (km s$^{-1}$) & No. of Cells & No. of Part. & t (s)\\
            \tableline

 1S      &70  &5 &0.0072&0.24&0.4&3900&28&54\,M  &2.1\,B & 21 \\
 2M      &35  &10&0.0036&0.48&0.8&3900&28&427\,M &17\,B & 17 \\
 3M      &35  &5 &0.0036&0.2&0.4&3900&32&427\,M &17\,B & 30 \\
 4M      &35  &5 &0.0013&0.2&0.2&7800&32&427\,M &17\,B & 17 \\
 5L      &17.5&5 &0.0018&0.1&0.4&3900&22&3.4\,B  &136\,B &12 \\
 3M$_{T}$    &35  &5.5 &0.0036&0.77&1.3&3900&63&427\,M &17\,B &12 \\
 2M$_{T}$    &35  &11 &0.0036&1.55&2.6&3900&67&427\,M &17\,B &12 \\
             \tableline
        \end{tabular}
\begin{tablenotes}
\item \textbf{Notes.} \small Given are median values over the entire domain for all the ratios of parameter per cell size $\Delta s$. $\delta_e = c /
\omega_{pe}$ is the skin depth, where $c$ is the speed of light, $\omega_{pe}$ is the plasma frequency, $\Delta t$ the time step, $r_g \approx v_{th,e} / \omega_{ce}$ the gyro
radius, where $\omega_{ce}$ is the electron gyro-frequency, $\lambda_D = v_{th}/\omega_{pe}$ is the Debye length and $v_{J}=J/(q n)$ is the electron
current velocity. Further listed are the number of cells and particles used
for each simulation and the simulated time $t$ in solar seconds. The runs with
the attribute T have twice the temperature compared to the other simulations
runs.
\end{tablenotes}  
\end{threeparttable}
\end{table*}

\section{Results and Discussion}
\label{sec:results}
There are several objectives with these experiments, but the most important one
is to establish whether any particle acceleration (i.e., production of
non-thermal particles) takes place in the experiments and if so, what the main
particle acceleration mechanism at work is. As we demonstrate and discuss
below, non-thermal particles are indeed being produced. Their energy
distributions are approximate power laws, with slopes similar to the slopes
inferred from observations of the bulk of sub-relativistic accelerated
electrons in magnetic reconnection events in the solar corona.

Most importantly, the presented simulations manifest the main electron
accelerator as being a systematic electric field, which evolves relatively
slowly, and which we therefore characterize as being essentially a `direct
current' (DC) electric field. This does not mean that the field is completely
stationary, nor that time evolution is unimportant. What it does mean is that
we can demonstrate that power-law distributions of electrons can be created
even by nearly stationary electric fields, as a result of basically geometric
factors, and without reliance on a recursive process in time. Such systematic
electric fields occur in our experiments in connection with the strong current
sheets that build up as a consequence of the imposed stress on the magnetic field.

\begin{figure*}
    \centering
    \includegraphics[width=0.8\linewidth]{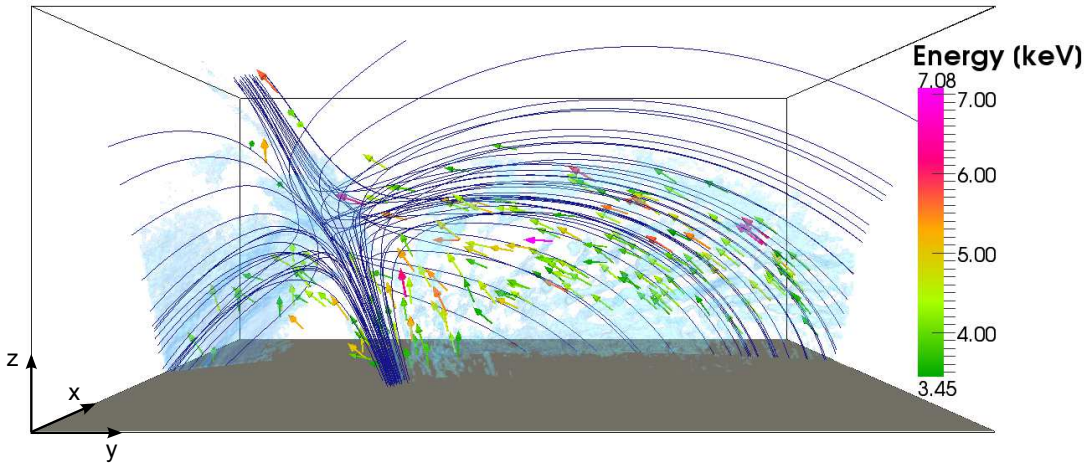}
    \caption{Random selection of non-thermal electrons (arrows with
    color indicating their energy) in the
current sheet (semitransparent light blue), which is located in the fan-plane,
together with the magnetic field (dark blue) for run\,3M. The spine of the
topological structure stretches into the left upper corner (outer spine) and
reaches down into the center of the dome-shaped fan-plane (inner spine).}
    \label{fig:eplane_jcontour_energyparticles2.eps}
\end{figure*}

In the beginning of the PIC simulations we generally observe a strong increase in the electric current
density and the electric field in the fan-plane, close to the location where
the relative inclinations between the magnetic field lines of the two
conductivity domains---the inner and outer fan-plane area---are largest, but also
close to the inner spine.
Due to the boundary conditions chosen, we see a motion of the inner spine
as well as a relative motion between the inner
and outer spines, where their relative distance increases with increasing stress and
decreases as the system relaxes, as discussed also in the corresponding MHD simulation article \citet{2013SoPh..284..467B}.
The photospheric boundary
motion triggers magnetic reconnection at the null-point by adding shear between
the inner and the outer magnetic field lines of the fan-surface, which causes stress
on the system that can only be reduced by dissipative processes, including
magnetic reconnection. The resulting current sheet after 8 seconds of solar time in simulation run\,3M is shown in Figure \ref{fig:eplane_jcontour_energyparticles2.eps},
which also shows the location of
the non-thermal high energetic electrons (arrows), located in the current sheet
of the fan-plane outlined by magnetic field lines, coinciding with the
location of an enhanced electric field and \textit{E$\parallel$$\hat{\textrm{B}}$}. The
semitransparent light blue surface depicts the highest electric current
density, which is located in the fan-plane. A
fraction
of the high-energetic electrons move through the null-point up
along the outer spine. Due to the increasing magnetic field strength with
increasing distance from the null,
these particles will essentially follow the magnetic field lines, and will, unless
reflected by a magnetic mirroring effect, likely impact
the solar surface where the strongly bent outer spine intersects with the
photosphere. The other (larger) fraction of non-thermal particles passes the null-point and
impacts the  photosphere on the north-west side of the fan-plane (to the left
in Figure \ref{fig:eplane_jcontour_energyparticles2.eps}). A comparison of this
impact region with the observations of this
event reveals several similarities, as discussed in Section \ref{sec:observations}. The
electrons in the fan-plane experiences a strong acceleration parallel to the magnetic field inside
the current sheet while approaching the null-point.

\subsection{The Occurrence of Electric Fields in MHD and PIC Experiments}
A systematic electric current develops in the fan-plane as a result of the photospheric
driver, which introduces a near discontinuity in direction of the magnetic field
in the two regions with different connectivity; outside and inside the fan-plane.
This occurs already in the MHD
experiment, where the electric current $\JJ$ is identically equal to
$\nabla\times\BB$ (disregarding constant factors), and where the corresponding
electric field in the local reference frame, here called the {\em diffusive electric field},
is given by an Ohm's law of the
type $\EE_\eta = \eta \JJ$, where $\eta$ is a numerical resistivity.

Apart from rapid fluctuations (on plasma frequency time scales) a similar
equality between $\nabla\times\BB$ and the electric current $\JJ$ must also hold in
the PIC experiment. Because the PIC simulations are collisionless one would
perhaps expect that the diffusive part of the electric field, initially
not inherited from the MHD simulation (only the advective, $-\uu\times\BB$, part is kept
in the PIC initial condition), would remain negligibly
small. The results of our experiment shows, however, that this is not the case.
Instead of remaining small, the diffusive electric field actually grows in magnitude,
although it generally remains much lower (after taken into account the modifications
as per Section \ref{sec:DCefield}) than in the MHD simulation from which the PIC
simulations were started.

On closer inspection, the reason for the growth of the electric field becomes
obvious: the charged particles that carry the electric current --- this is
primarily the electrons --- cannot go wherever they want by following the
slightest whim of a tiny electric field (in the frame of reference moving with
the current).  Instead, they are in general forced to follow magnetic field
lines.  But even along field lines, the charged particle motions are not
unhindered; already modest (local or global) increases of the magnetic field can
force a charged particle to become reflected.  In addition, systematic or
fluctuating cross-field components of the electric field may cause charged
particles to drift away from its initial field line, especially in low magnetic
field strength regions, and in regions with significant shear of magnetic field
lines; i.e., precisely in regions with significant net electric currents.

In particular, it is clear from the very definition of the concept that
magnetic reconnection causes a continuously on-going change of magnetic
connectivity, and that this change occurs exactly in the place where the
largest electric current needs to be maintained.

\begin{figure}
    \centering
    \includegraphics[width=1.0\linewidth]{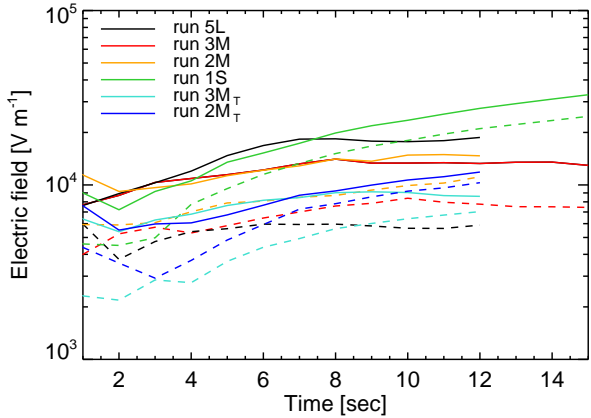}
    \caption{Average magnitude of the largest total electric field (solid) and
    \textit{E$\parallel$$\hat{\textrm{B}}$} (dashed) calculated on a cut-out around the
    null-point with dimensions $13\times21\times16$\,Mm over the number of
    grid points covering 1\,Mm for runs 2M, 3M, and 5L and the higher temperature runs 2M$_{T}$ and 3M$_{T}$.}
    \label{fig:eparab_plot.ps}
\end{figure}

The component of the electric field along \textit{B}, \textit{E$\parallel$$\hat{\textrm{B}}$},
corresponds closely to the diffusive electric field in MHD when considering the
corona. \textit{E$\parallel$$\hat{\textrm{B}}$} is plotted together with
the magnitude of \textit{E} in V\,m$^{-1}$ for runs 1S, 2M, 3M, 5L, 2M$_{T}$, and
3M$_{T}$ in Figure \ref{fig:eparab_plot.ps}.
In the snapshot from the MHD run used for the PIC simulations, we find
diffusive electric field values of about
29\,V m$^{-1}$ and a total electric field of 1666\,V m$^{-1}$. A
comparison of the locations of high diffusive electric fields in the MHD with
regions of high \textit{E$\parallel$$\hat{\textrm{B}}$} in the PIC simulations show that the peak values can be found in both cases in the current sheet, while
when comparing the advective electric fields, it reveals that, unlike in MHD case,
the PIC advective electric fields peak again in the current sheet. In the MHD
case, this
quantity is high in most of the domain around the null-point, while in the PIC
case it is only high inside the current sheet.
Both differences (magnitude and peak location of the advective electric
field) are presumably a consequence of lowering the elementary charge, imposed
by the modifications. The lower the charge, the higher the speeds of the
(lightest) particles have to be in order to maintain the same electric current,
set by the magnetic field geometry. Therefore, the diffusive as well as the
advective electric field adapt to higher values in the PIC simulations,
particularly in the current sheet. As previously mentioned, our modifications
of the constants of nature prevent us from directly comparing these electric
field values to the values obtained in the MHD simulations. But, as shown in
the next section, we are able to make a qualified guess as to what the PIC electric
fields would be in the real solar case, which we further compare to the MHD
electric fields. While most medium-resolution runs (cf. Table \ref{tab:SimulationRuns})
show comparable electric fields (Figure \ref{fig:eparab_plot.ps}), with
differences of a factor of 2--3, run 1S is under-resolved, and clearly
disagrees with all other simulation runs.

Charged particles moving in a realistic (non-smooth) electro-magnetic field
effectively experience a `resistance', and a non-negligible,
systematic electric field is needed to maintain the electric current $\JJ$
consistent with $\nabla\times\BB$, as required by the Maxwell equations. Below
we demonstrate that, at least in our numerical experiment, it is this
systematic electric field that is mainly responsible for the particle
acceleration.  We then argue that, since the same phenomena must occur in the
real solar case, a similar particle acceleration mechanism must be at work there.

\subsection{The DC Electric Field as the Particle Accelerator\label{sec:DCefield}}
The systematic electric field which develops across the fan-plane peaks where
the electric current density is largest. This is where particles are mainly being
accelerated and where the major contribution to the power-law energy distribution
population presented in Section \ref{sec:distr} comes from. Its origin is the
tendency for an imbalance between the current density and the curl of the magnetic
field, which occurs as a consequence of the dissipative reconnection processes. As
magnetic field lines are reconnected, the charged particles that flow along them
become `misdirected' and need to be replaced by the acceleration of new
particles that are now, instead of the previous ones, situated correctly with
respect to the magnetic field and its curl.

\begin{figure*}[!ht]
    \centering
    \includegraphics[width=0.75\linewidth]{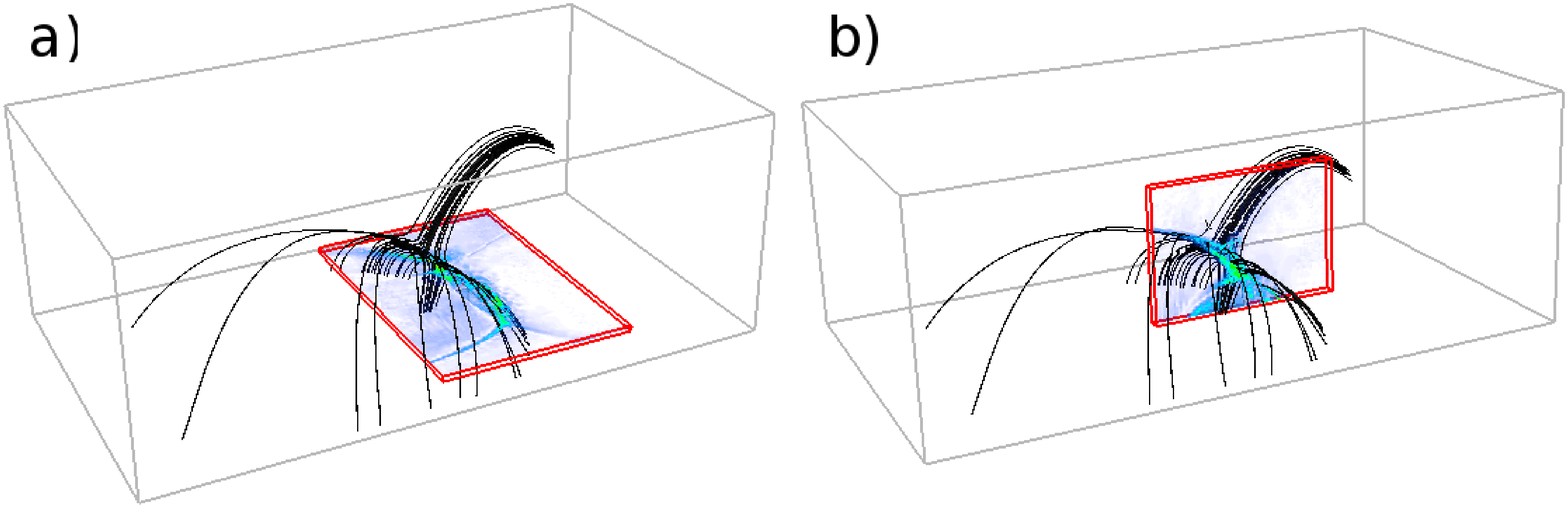}   
    \includegraphics[width=0.95\linewidth]{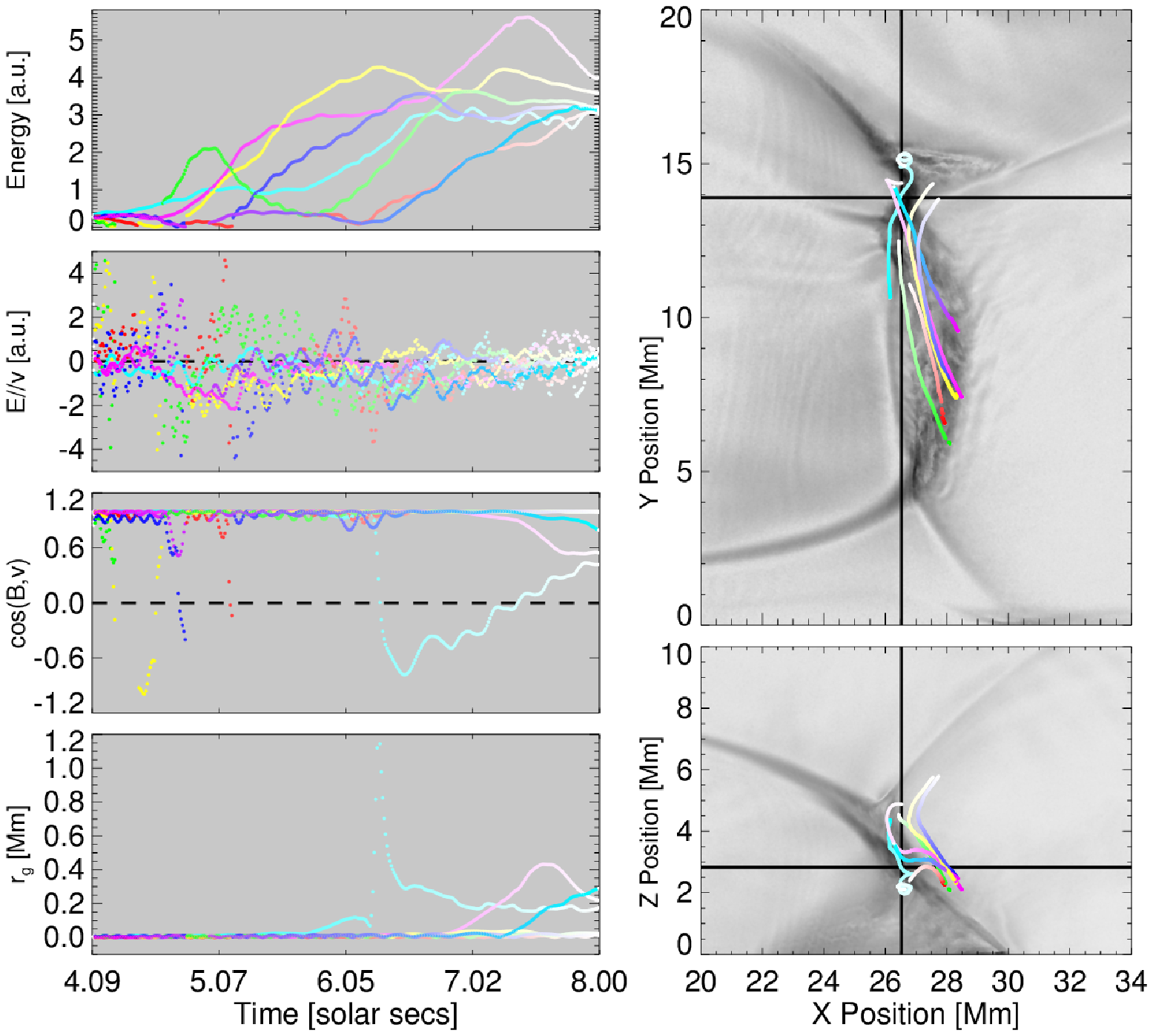}
    \caption{Seven randomly chosen accelerated electrons (colored lines) from the power-law energy tail of run 3M. The background gray-scale images to the right are the sum over electric current density slices. See the text for detailed information.}
    \label{fig:trace_plot_win.eps}
\end{figure*}

Tracing particles with kinetic energies in the high-energy tail and interpolating the
fields to the particle position in time-steps of 0.003 seconds for 5 solar seconds in
run 3M, the direct correlation between a continuous energy gain for electrons
and a negative electric field component in the direction of motion is apparent.
These non-thermal particles are primarily found in regions close to, or inside,
the current sheet of the fan-plane, where the electric current density is highest. In Figure \ref{fig:trace_plot_win.eps}, seven representative
non-thermal particles that gain energy during approximately 4 solar seconds are shown using
different colors. Plotted to the left are their energy, the electric
field component pointing in the direction of the particle velocity
\textit{E$\parallel$v}, the cosine of the pitch angle cos$(B,v)$, and the gyro-radius
$r_g$. An indication of the particle speeds may be obtained by observing the
paling
of the line color in their trajectories to the right which are superposed on
the images of the sum over electric current density slices. The upper panel shows the sum of $xy$-slices from a
height of 2.5--2.9\,Mm illustrated in (a) (see top of figure), while the
lower panel displays $xz$-slices averaged over 12.6--13.4\,Mm in the
$y$-direction as shown in (b) of Figure \ref{fig:trace_plot_win.eps}. Both background images are taken at time \textit{t} = 6 seconds and raised to the
power 0.5 to enhance the visibility of the fine structures. It is
essential to mention that these images only change slightly during the
presented time interval. The null-point is
located at [26.5, 13.9, 2.8]\,Mm, illustrated by a cross-hair. Note, the particle trajectories are marginally displaced, due to their projection onto the current density planes.

\begin{figure*}[!ht]
    \centering
    \includegraphics[width=0.95\linewidth]{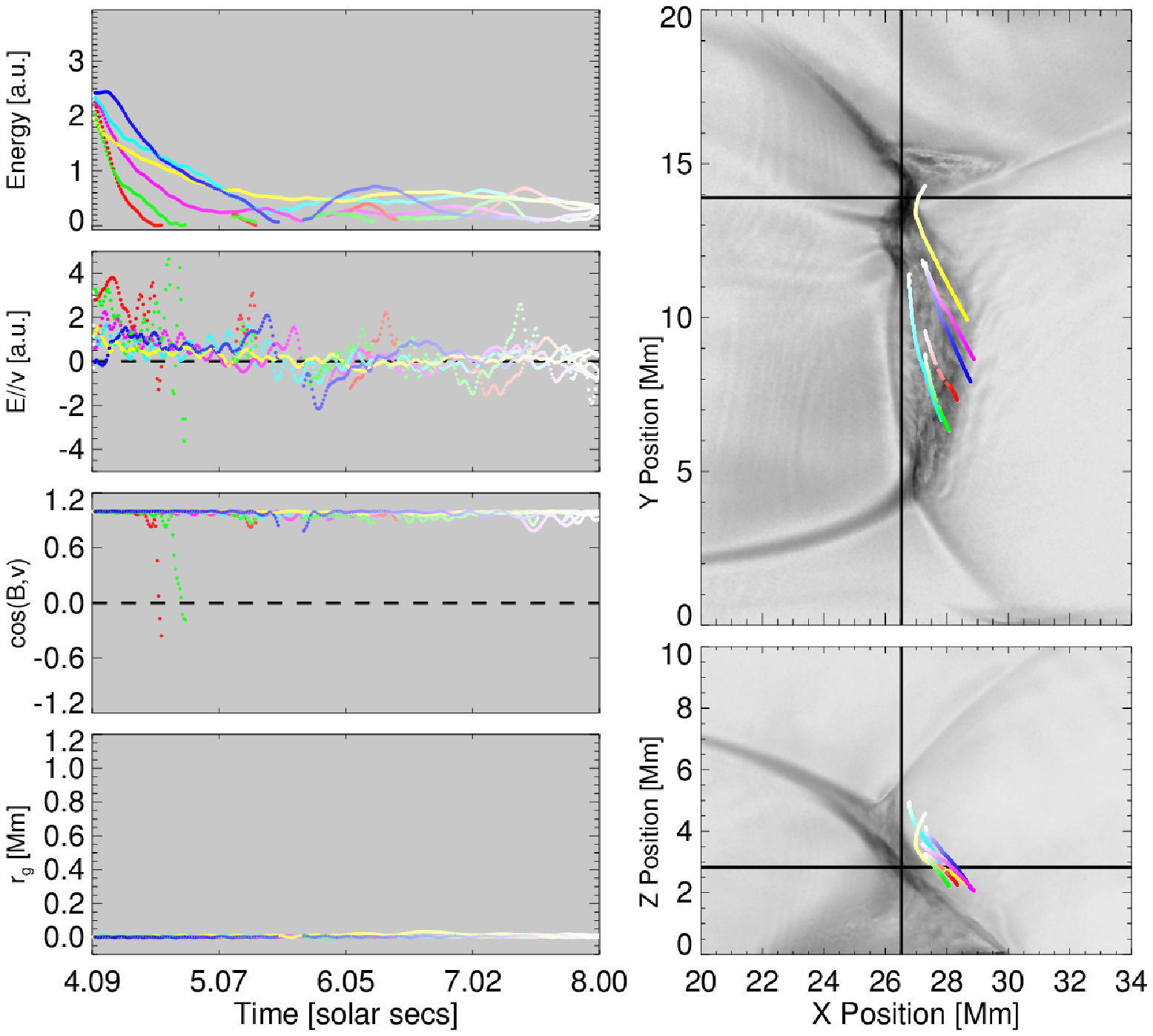}
    \caption{Analog to Figure \ref{fig:trace_plot_win.eps}, but for randomly chosen electrons from the power-law energy tail of run 3M, which lose energy.}
    \label{fig:trace_plot_lose.eps}
\end{figure*}

Similar plots, but for
particles that lose energy, are shown in Figure
\ref{fig:trace_plot_lose.eps}. Electrons starting from the right-hand side of
the current sheet feel first a very diffuse and rapidly changing electric
field, due to a fragmentation of the current sheet, which first develops there.
This behavior is most evident when looking at the green energy bump around time
5 seconds in the first panel of Figure \ref{fig:trace_plot_win.eps}. In such
fluctuating regions, there is a constant
competition between the dominance of the perpendicular electron movement
in-between the strong \textit{E$\parallel$B} patches and the parallel motion inside an
\textit{E$\parallel$B} region. But once a particle is inside a current filament the
velocity is mostly directed oppositely to the electric field, and electrons
then experience a rapid acceleration, due to which the perpendicular energy
relative to the magnetic field becomes negligible and the
electrons move almost parallel to the magnetic field (cf.\ the cos(pitch angles) in
Figure \ref{fig:trace_plot_win.eps} and \ref{fig:trace_plot_lose.eps}).
The gyro-radius is permanently very small (on the order of 3 -- 8\,km), meaning
that electrons tightly follow the magnetic field lines. When particles
approach the null-point (marked by the cross-hair in the graphics), their gyro
radius naturally grows, as $r_{g} \propto v_{\perp} \cdot B^{-1}$. At the same
time, the magnetic field is weak enough that the electrons are no longer strongly
confined to the magnetic field lines, and they are instead directly accelerated by
the electric field, rather independent of their orientation relative to the magnetic
field. Particles entering the region close to the strongest current sheet may
also experience a modest growth in gyro-radius, as the magnetic field there is
rather weak.

Small-scale regular oscillations which can especially be seen in the
\textit{E$\parallel$v} and in the cos(pitch angle) in the panels of Figures
\ref{fig:trace_plot_win.eps} and \ref{fig:trace_plot_lose.eps} occur at a frequency of about
7.65\,s$^{-1}$, which is of the order of the plasma frequency.

We provide in Figure \ref{fig:eparab_plot.ps} as well as in Figures
\ref{fig:trace_plot_win.eps} and \ref{fig:trace_plot_lose.eps} electric field
values which are affected by our modifications of the constants of nature. We
therefore need to address how to transform these values back, in order to make
them comparable to the MHD simulation results as well as to observations.

For this purpose we start out with the force that an electron with charge $q$
and mass $m_e$ experiences when moving in an electric field $E$, assuming that
the DC electric field is the main particle accelerator $F=q\,E=m_e\,a$. We further
neglect the displacement current since the accelerating electric field---ignoring
small-scale turbulence---is persistent. At time $t$ and after having covered a
distance $L=\frac{1}{2}\, a\, t^2$ the electron has a velocity $v$ of
\begin{equation}
  v = a\,t =(2 L a)^{1/2} = \left(\frac{2 L q E}{m_e}\right)^{1/2}.
\end{equation}
The electric current has to balance the magnetic field through Amp\`ere's law
\begin{equation}
  J = \mu_0^{-1} \nabla \times B \approx \mu_0^{-1} \frac{\Delta B}{\Delta L}\,,
\end{equation}
where $\Delta B$ is the typical change in the magnetic field across
the current sheet and $\Delta L$ is the thickness of the current sheet. The
current is generated by moving the electrons
\begin{equation}\label{equ:electric current}
J = n v q = n \left(\frac{2 L q^3 E}{m_e}\right)^{1/2},
\end{equation}
where $n$ is the particle density. Using the above two expressions for $J$, we
obtain an equation for the total
energy gain of a single particle due to the electric field
\begin{equation}
  q\,E\,L = \mu_0^{-2} \left(\frac{\Delta B}{\Delta L}\right)^2 \frac{m_e}{2 n^2 q^2} .\label{equ:efield}
\end{equation}
Given that this acceleration only happens inside the current sheet the
expression can be interpreted as the maximal acceleration a single electron can
obtain, if it moves continuously inside the current sheet over the distance
$L$.

To relate Equation\,(\ref{equ:efield}) to our simulations and real observations,
we need the scaling of $\Delta B$ and the current sheet thickness $\Delta L$.
$\Delta L$ is determined by stability constraints, as instabilities diffuse the
current away, in case it shrinks below a certain thickness $\Delta L$.
There have been several studies in 2D and 2.5D on the current sheet thickness
concluding that $\Delta L$ is comparable to the size of the diffusion region, as noted in
laboratory experiments by \citet{2008GeoRL..3513106J}, as well as in PIC
simulations by \citet{2001JGR...106.3721H}. But these magnetic geometries are
not directly comparable to our case, in which most of the diffusion takes place
in the current sheet rather than around an idealized magnetic X-point geometry.
In fact, very little is known about the thickness of 3D fan-plane current
sheets. We therefore cover here the most probable cases.
We assume the smallest length scale over which a coherent large-scale plane of
current can be maintained is either the electron gyro-radius $r_g$ in
the magnetic field or the electron skin depth:
\begin{align}
 \Delta L &\approx \frac{m_e v_{th,e}}{q B} \label{equ:DeltaL} & \textrm{(gyro-radius)} \\
 \Delta L &\approx \left(\frac{m_e}{\mu_0 n q^2}\right)^{1/2} \label{equ:DeltaL2} & \textrm{(skin depth)}
\end{align}
(if $\Delta L$ is instead of the order of the ion gyro-radius it would be
larger than the estimate in Equation\,(\ref{equ:DeltaL}) by a factor $(m_p/m_e)^{1/2}$,
which is about a factor of 4 in our case).
If we rewrite $\Delta B$ as a fractional change in the magnetic field $\Delta B = \epsilon_B B$, and express $B$ and
$\Delta B$ in terms of the Alfv\'en speed $v_A$ in the plasma
\begin{equation}
v_A^2 = \frac{B^2}{\mu_0 n m_p} = \frac{\Delta B^2}{\epsilon_B^2 \mu_0 n m_p}
\end{equation}
two elegant expressions emerge for Equation\,(\ref{equ:efield}), defining the
maximal electron energy generated by the DC acceleration
\begin{align}
q\,E\,L &= \epsilon_B^2 \frac{E_{A}^2}{E_{th,e}}\,,  \label{equ:EqL} & \textrm{(}\Delta L\sim\textrm{gyro-radius)} \\
q\,E\,L &= \epsilon_B^2 E_{A}\,,  \label{equ:EqL2} & \textrm{(}\Delta L\sim\textrm{skin depth)}
\end{align}
where $E_{A} = \frac{1}{2} m_p v_A^2$ is the ``kinetic Alfv\'en energy'', which
is needed to move information in the system, and $E_{th,e} = \frac{1}{2} m_e
v_{th,e}^2$ is the thermal energy of the electrons. If the current sheet thickness
is related to the gyro-radius, then the higher the temperature the smaller the acceleration of
an individual particle. A higher temperature is also reflected in a larger gyro
radius, but the total energy available for acceleration induced by Amp\`ere's
law is the same, and hence there must be more, but lower energy, particles in a
thicker current sheet. On the other hand, if the current sheet thickness is
related to the electron skin depth the maximum energy should be independent of
the temperature. Note that the right-hand side of Equations\,(\ref{equ:EqL}) and
(\ref{equ:EqL2}) only contains macroscopic fluid parameters. Equations (\ref{equ:EqL}) and (\ref{equ:EqL2}) provide an estimate of the electric
field and its dependency on the charge per particle; the
charge times the electric field magnitude is to lowest order a constant and
hence our modifications in charge $q_{mod}$ are reflected in a $q_{mod}$ times too large
electric field, where typically $q_{mod} \approx 2\e 6$. Equations similar to
\ref{equ:EqL} or \ref{equ:EqL2} are what we expect to be able to test in the future.
As for now the numerical resolution is not sufficient for an experiment to
resolve typical gyro radii with enough grid cells, while
simultaneously resolving the large-scale plasma.

In the current experiments, the current sheet thickness is essentially
determined by the grid spacing, and is generally a few times $\Delta s$;
about an order of magnitude larger than the typical electron gyro radii,
and about half an order of magnitude larger than the typical proton gyro
radii.  If, at the same elementary charge per particle $q$, we were able to
increase the numerical resolution in order to resolve the gyro radii and the current sheet became correspondingly
thinner, then according to Equation\,(\ref{equ:efield}) the maximum energy gain would
increase with the square of the $\Delta L$ factor, so with 1 -- 2 orders of
magnitude.

We conclude that a conservative estimate of the current sheet electric field
in the Sun would be smaller by at most the factor (about $2\e{6}$) by which
the elementary charge per particle has been reduced, and it could possibly be
1 -- 2 orders of magnitude larger.  Taking 4000\,V\,m$^{-1}$ as a typical magnitude
of the electric field in our experiments, we thus come up with a conservative estimate
on the order of 2\,mV\,m$^{-1}$ for the solar electric field in a situation analogous
to the one we model (which is {\em not} a flaring situation). Using a less conservative
estimate based on the argument of numerical resolution, the electric field is on the order of 20 -- 200\,mV\,m$^{-1}$. In comparison, electric fields
\textit{during} solar flares are inferred to be on the order of thousands of V\,m$^{-1}$ \citep{2002ApJ...565.1335Q}.

\subsection{Magnetic Field Geometry versus Power-law Distribution\label{sec:Geometry}}
\begin{figure}
    \centering
    \includegraphics[width=0.9\linewidth]{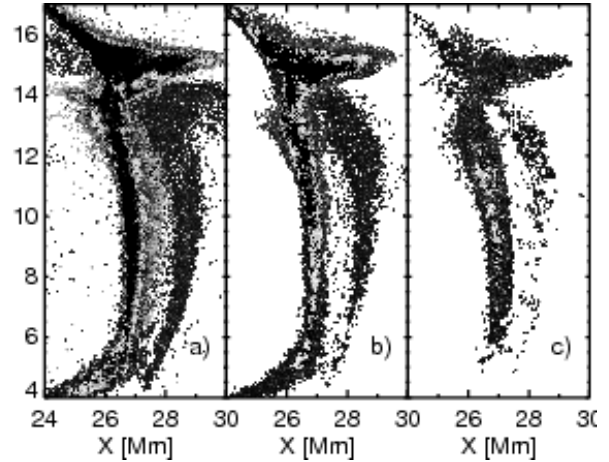}
    \caption{Run 5L: power-law tail electric current density ($J$) contributions from
particles within the energy range of (a) 4.3\e{-22} $< e <$ 1.4\e{-21}, (b)
1.4\e{-21} $< e <$ 4.4\e{-21}, and (c) $e >$ 4.4\e{-21} in (a.u.).}
    \label{fig:current_contrib}
\end{figure}

\begin{figure}
    \centering
    \includegraphics[width=0.9\linewidth]{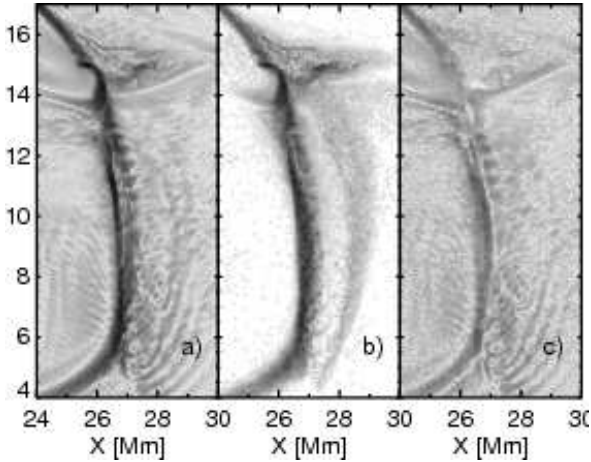}
    \caption{Electric current density arising from (a) bulk flow, (b) the power-law
tail particles, and (c) the thermal particles in run 5L.}
    \label{fig:current_sheet_contrib}
\end{figure}

\begin{figure}
    \centering
    \includegraphics[width=0.9\linewidth]{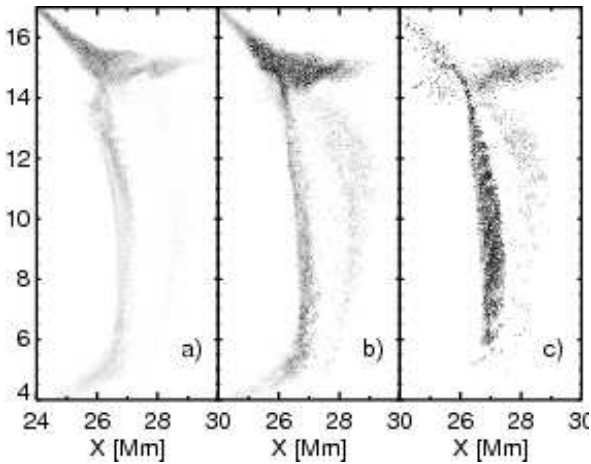}
    \caption{Run 5L: power-law tail energy ($e$) contributions from particles
within the energy range of (a) 4.3\e{-22} $< e <$ 1.4\e{-21}, (b)
1.4\e{-21} $< e <$ 4.4\e{-21}, and (c) $e >$ 4.4\e{-21} in (a.u.).}
    \label{fig:energy_contrib}
\end{figure}

As the current sheet channels make up a very small fraction of the domain, most
particles will not exhibit the correct angle to exactly pass through an
acceleration channel, but will instead be deflected by the magnetic field,
ending up outside the current sheet without undergoing a continued
acceleration. Hence, most particles are \emph{not} continuously accelerated. The
electric current itself is mainly carried by the lowermost part of the
power-law distribution, as illustrated in Figure \ref{fig:current_contrib},
which shows the contributions from the particles to the electric current
density from three equally large regions of log(energy) of the power-law tail
of run 5L. The particles of the lowest bin are most probably in constant
exchange with the thermal particle distribution. Figure
\ref{fig:current_sheet_contrib} sets the power-law tail in relation to the bulk
flow (average velocity of a given species inside a cell) and the thermal particles. The power-law tail dominates over the thermal
contribution with respect to contributions to the electric current density.
About 0.2\% of all electrons in the computational box make up the non-thermal
electrons high-energy tail of the distribution. 5\% of the total electron
energy is carried by the power-law tail particles, while most of this energy is
in the most energetic particles of the power-law tail, as illustrated in Figure
\ref{fig:energy_contrib}. But, if we only consider particles with a negative
vertical velocity, thus moving toward the bottom of the box, and additionally
reside in a zone on the lower quarter of the box of 1.9\e{11}\,km$^3$ (from
approximately 0.175 -- 1.750\,Mm above the bottom boundary), the
energy share coming from the power-law tail amounts to over 50\% and we find
more than 5\% of all particles in the high-energy tail population in run 5L, while the
total number of electrons in this cut-out is about 1.52\e{7}.
These are the particles that on the real sun would be decelerated in the
chromosphere, leaving an imprint in the form of observable bremsstrahlung emission.

\subsection{Comparison with Observations\label{sec:observations}}

\begin{figure}[ht]
    \centering
    \includegraphics[width=0.9\linewidth]{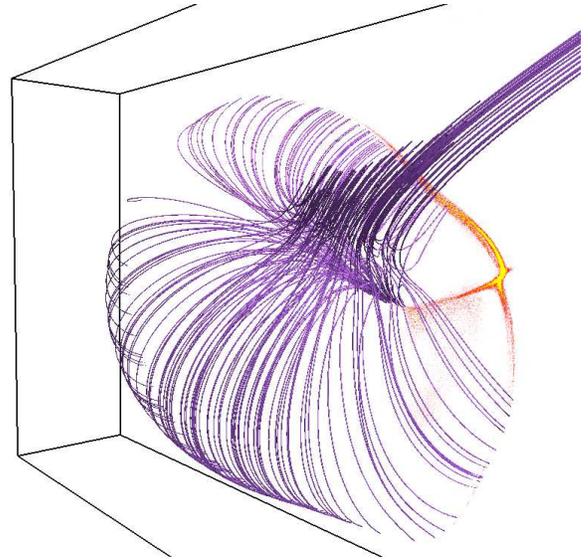}
    \caption{Impact area of mainly non-thermal electrons from run 5L. Shown is the electron energy
    (from orange increasing to yellow) added up at the lower boundary over a
period of 5 seconds. Additionally, magnetic field lines (purple) passing close to
the null-point are plotted.}
    \label{fig:impact_area_snap.eps}
\end{figure}

Considering the observations of this particular reconnection event, in addition to the
bright ribbons observed in H$\alpha$ and UV at the intersection of the fan and
the chromosphere \citep{2009ApJ...700..559M}, the chromospheric foot-points of
the interchange reconnection region show also soft and hard X-ray signatures during the impulsive phase with a peak intensity slightly northward of the null-point, as shown at the top of Figure 5 in \citet{2012A&A...547A..52R}. This coincides well with the impact region of the power-law electrons in our
simulation which travel along the fan-plane and finally hit the
lower boundary. Figure \ref{fig:impact_area_snap.eps} shows the non-thermal
electron energies in the simulation at their impact regions on the lower boundary of the box
accumulated over \textit{t} = 4 -- 9 seconds.
The small difference in the location of the peak intensity compared to
observations \citep{2009ApJ...700..559M,2012A&A...547A..52R} can at least partly
be explained by the driving pattern, which is of course only an approximation
to the real photospheric boundary motion. The second reason may be the overall
magnetic field at the start of the simulation, which is an outcome of an MHD
simulation, which was again initialized from a
potential field extrapolation.

However, our electron energies are clearly lower than what is needed for the
observed emission spectra to be produced. Hence, it is important to emphasize that
we do not model the observed flare event, but rather the pre-flare phase. One
reason is that we use an MHD state taken at a time well before the
flare event. Another reason is, as shown in \citet{2013SoPh..284..467B}, that the
boundary driver in the MHD simulations, representing the observed horizontal
magnetic field motion in the active region, does not provide
enough shear and stress to the system to result in an abrupt energy release.
Despite that, we expect the qualitative nature of the acceleration mechanism
to be the same in the flare phase as in its pre-phase, since the \textit{overall} magnetic morphology during a flaring process does not significantly change, but displays a larger current and hence a larger parallel electric field is required.

\subsection{The Energy Distribution and the Influence of
the Modifications}\label{sec:distr}
\begin{figure}
    \centering
\includegraphics[width=0.7\linewidth]{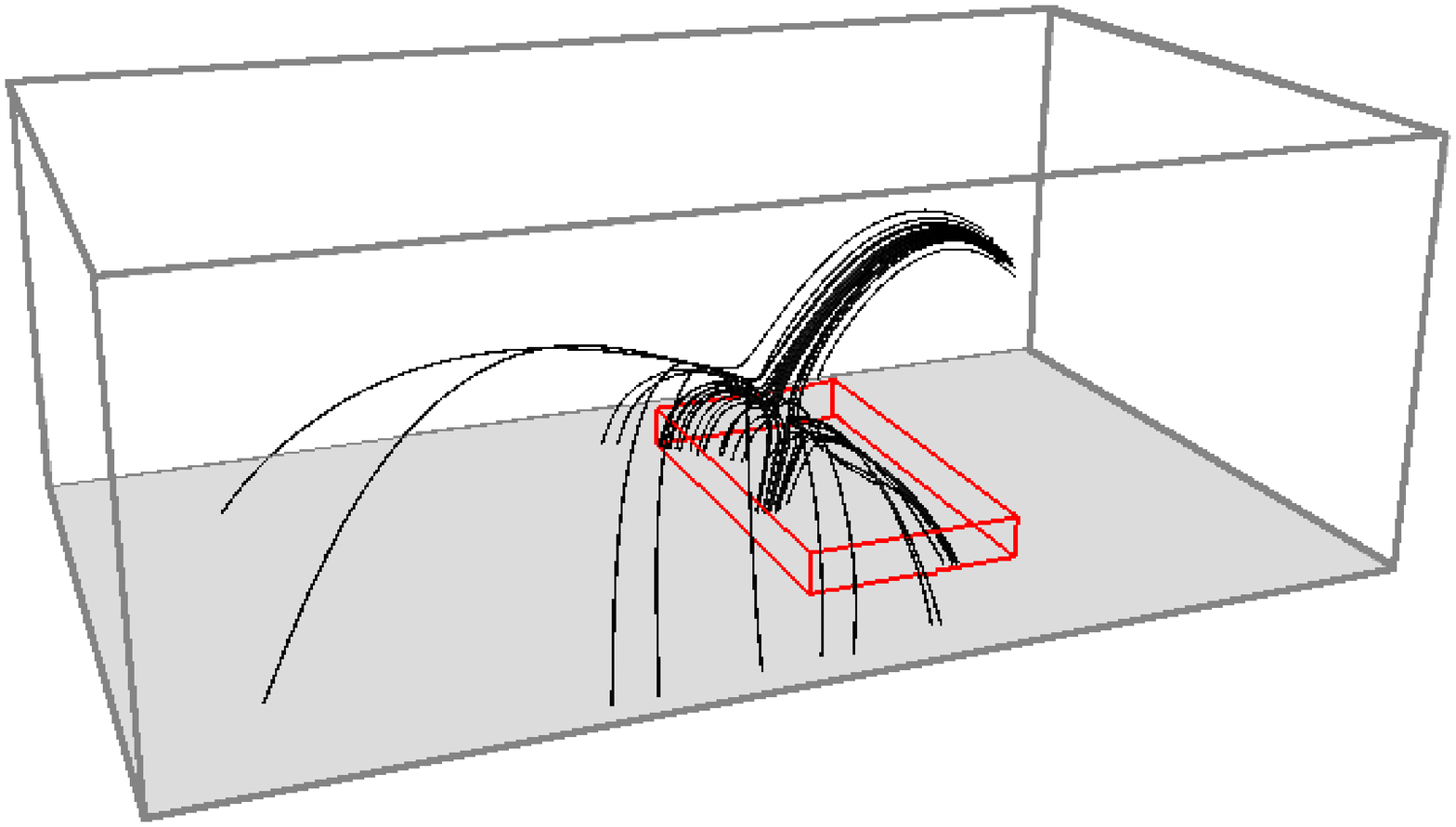}
    \label{fig:histogram3_cutout}
    \includegraphics[width=1.0\linewidth]{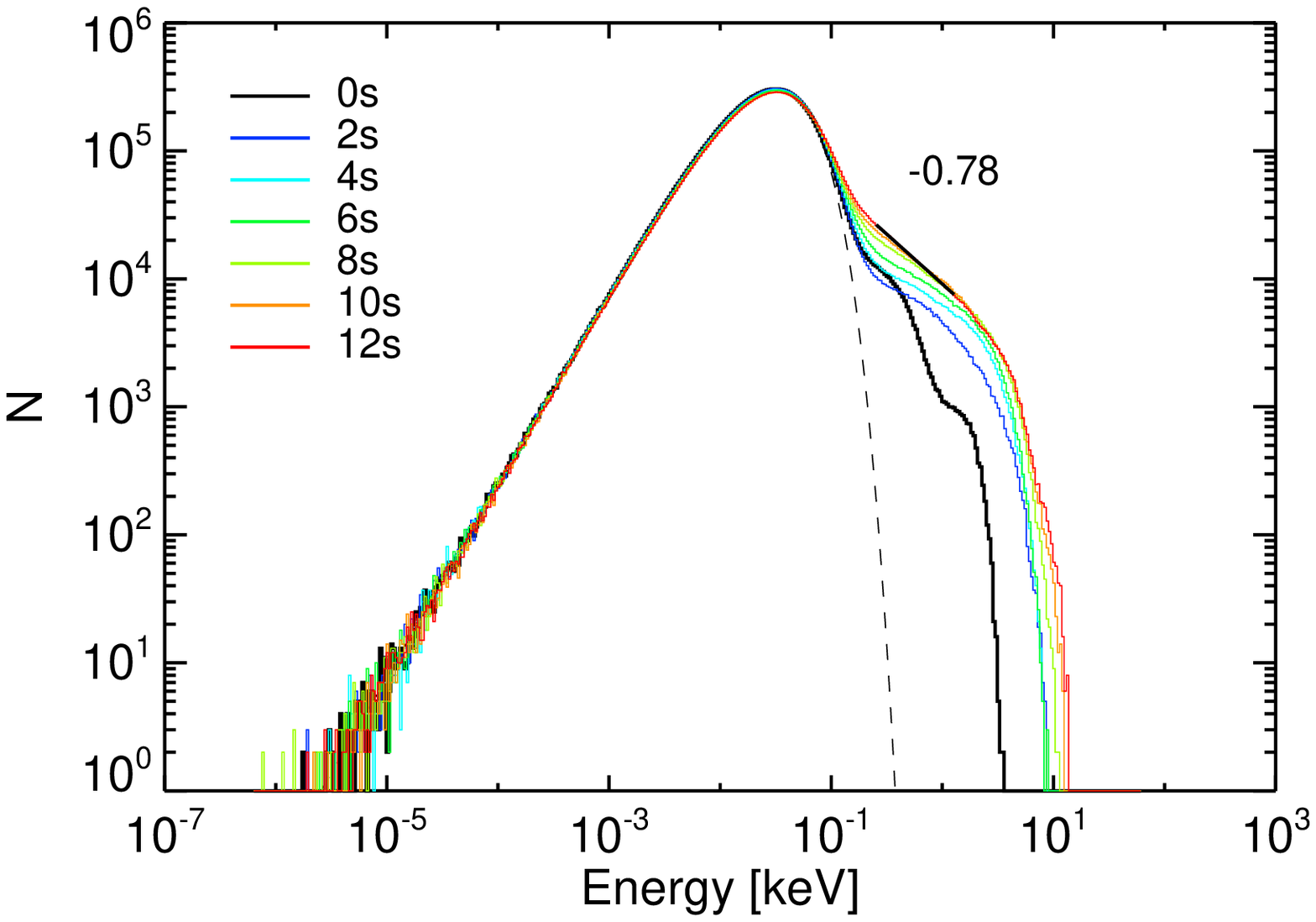}
    \caption{Energy distribution of run 5L covering 12 solar seconds from the beginning of the simulation, obtained from 1.52\e{7} electrons in a volume of
1.9\e{11}\,km$^3$ shown in the illustration above the plot. The dashed line represents a Maxwell--Boltzmann distribution fit.}
    \label{fig:histo}
\end{figure}

Looking at the energy histogram of the same particles considered in
Section \ref{sec:Geometry}; downward moving in a cut-out of the lower part of
the simulation box (see top illustration in Figure \ref{fig:histo}), we find a
Maxwell--Boltzmann distribution combined with a $dN/d\ln E  = E dN/dE$ power-law
index of about -0.78, corresponding to a $dN/dE$ distribution power-law index
of -1.78 (see Figure \ref{fig:histo}). A power-law index of -1.78 implies that
the electric current resulting from the power-law population is mainly carried
by the low-energy electrons, while the kinetic energy is mainly carried by the highest
energy constituents, visualized in Figure \ref{fig:energy_contrib}. The color
code in Figure \ref{fig:histo} shows the temporal evolution of the energy
distribution function for downward-moving particles in the cut-out. The
power-law index for the full simulation box is similar. Figure \ref{fig:histo}
further shows that the tail slope rapidly converges toward the power-law index
of about -1.78, indicating an impulsive acceleration of electrons. The weak,
apparently non-thermal tail present in the distribution function from the very
beginning (black solid line), does not share origin with the power-law tail that
is created dynamically at later times. It arises due to the electric current in
the MHD current sheet (the initial electron velocities are drawn randomly from
a Maxwellian distribution, shifted with a systematic velocity to maintain the
local electric current density). It is entirely due to the high values of the
drift velocity necessitated by our rescaled units, but does not influence the
later creation of accelerated particles; they appear also in tests where the
electrons are not given a systematic initial velocity.

\begin{figure}[ht]
    \centering
    \includegraphics[width=1.0\linewidth]{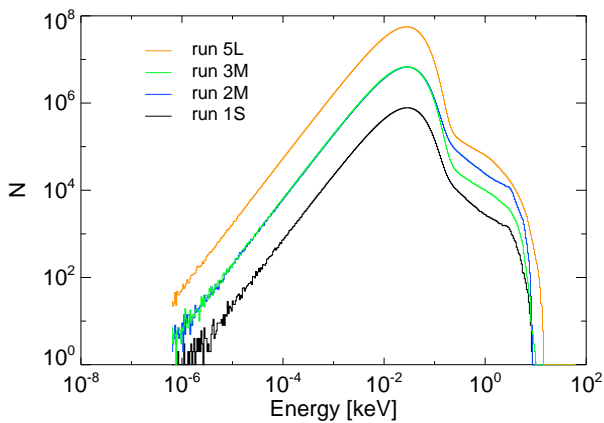}
    \caption{Energy histogram for all electrons in the full box for different simulation runs at time \textit{t}\,=\,12\,s.}
    \label{fig:energyhist_compare}
\end{figure}

\begin{figure}[ht]
    \centering
    \includegraphics[width=1.0\linewidth]{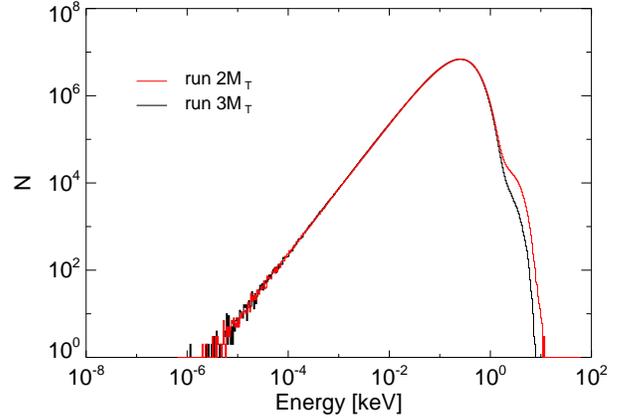}
    \caption{Energy histogram for all electrons in the full box at time \textit{t}\,=\,12\,s for the simulation runs with double the temperature of the runs presented in Figure \ref{fig:energyhist_compare}.}
    \label{fig:energyhist_compare_new_cutout}
\end{figure}

A comparison of the different electron energy distributions from the simulation
runs listed in Table \ref{tab:SimulationRuns}, as previously mentioned in Section \ref{sec:validation}, may be found in Figures
\ref{fig:energyhist_compare} and \ref{fig:energyhist_compare_new_cutout}. For these plots, all electrons in the
full simulation domain are used. The histograms are not normalized by the
number of particles in order to allow for a better power-law index comparison
between the different runs. The thermal distribution is the same for
all simulations having initially the same temperature profile and it is stable
over at least 12 solar seconds, as shown as an example in run 5L (Figure
\ref{fig:histo}). The non-thermal
energy tail part quickly approaches a power-law index of about $-1.78$ for all
runs, independently of their resolution. The power law itself is mainly a
consequence of the DC systematic electric field particle acceleration and
together with the cutoff of the energy histogram presumably a result of the 
available electric potential difference in the system, which is
determined by the current sheet thickness. The thickness, on the other hand, is
controlled by the magnetic
field geometry and its evolution.

For simulations with initially a higher temperature, see Figure
\ref{fig:energyhist_compare_new_cutout}, the non-thermal part of the energy
distribution drowns in the thermal part.
With a plasma beta in the corona
$<< 1$, a change in the temperature is not expected to significantly change the
overall electric field (also see Figure \ref{fig:eparab_plot.ps}) and
thereby the maximum energy that can be gained by a particle.

Observational support for the power-law index received from our simulations can be found in, e.g., \citet{2007ApJ...663L.109K}. There have also been a number of test particle investigations of current
sheets, such as
\citet{2006A&A...449..749T}, \citet{2005SoPh..226...73W} and \citet{2005SSRv..121..165Z} finding
similar electron power-law indices --- and even slightly harder. But such
comparisons are dangerous, as these these studies were conducted using the test
particle approach and the latter additionally assumed a simplified 3D magnetic
and electric field configuration, presumably having a significant influence on
the power-law index as partially studied by \citet{2005SSRv..121..165Z}.

\section{Conclusions}
\label{sec:conclusions}
We presented a first of a kind PIC study of a realistic AR topology.
In this study of a pre-flare 3D reconnection region near a 3D magnetic null
point in the solar corona, the main electron
acceleration mechanism has been shown to be the parallel systematic, almost stationary electric field (`DC' electric field),
building up as a consequence of the magnetic reconnection and dissipation
close to the null-point and in the fan-plane current sheet of this well-defined magnetic topology. We
estimate from the simulation results the average electric field in this
quiescent region to be on the order of 20 -- 200\,mV m$^{-1}$ in the Sun, which is
significantly smaller (2 -- 3 orders of magnitude) than what has been found in the corresponding MHD simulations
\citep{2013SoPh..284..467B}. However, this electric field is still sufficient to create a noticeable
non-thermal electron power-law tail with power-law indices on the order of what is observed during solar flares \citep[e.g.][]{2007ApJ...663L.109K}. We also present a discussion of the expected
dependency of the electric field on the modified elementary charge per particle;
showing that the product $E\cdot q$ is
expected to be conserved to lowest order. The verification of this relation must be
left for future studies due to the current restrictions on computational
resources.

The different modification options given in this article serve to
reduce the ratio between micro and macro-scales, which is required in order to overcome numerical constraints when simulating realistic physical scales of active regions. The main parameter change
used to increase micro-scales is a reduction of the elementary charge per
particle, essentially equivalent to a reduction of the particle density. As part of the model validation, we found the electric field values and the non-thermal energy distribution to be only weakly dependent on the applied modifications of the constants of nature. In spite of the modifications of the constants of nature necessary to make this
experiment possible we expect that the physical processes in the experiment are
qualitatively similar to those in the real Sun, and we therefore see these kinds of studies as a very valuable tool for studying the coupling between kinetic and MHD scales in semirealistic models.

From studies of the locations of the
non-thermal electrons and of their acceleration paths we conclude that the magnetic
field geometry and its temporal evolution are likely to be the main factors
controlling the power-law index measured in this experiment.
We further show that in the lower part of the computational box the
electron energy is predominantly in the
non-thermal component, with a particle impact area that correlates well with
the observations by \textit{TRACE}, \textit{SOHO} and \textit{RHESSI} \citep{2009ApJ...700..559M,2012A&A...547A..52R}.

In the future, we hope to strengthen our findings with even higher resolution
simulations, and we additionally plan to study the temporal dependence of the
power-law index by employing longer simulation runs, which would at the same time
provide an opportunity to study ion acceleration.

\section*{Acknowledgments}
\label{sec:acknowledgments}
Special thanks go to Jacob Trier Frederiksen and Klaus Galsgaard for very valuable discussions
and to Guillaume Aulanier and Sophie Masson for sharing their MHD data with us.

G.B. was supported by the Niels Bohr International Academy and the SOLAIRE
Research Training Network of the European Commission (MRTN-CT-2006-035484).
T.H. is supported by the Centre for Star and Planet Formation which is financed
by the Danish National Science Foundation.
The results in this paper have been achieved mainly using the PRACE and John von
Neumann Institute for Computing Research Infrastructure resource JUGENE/JUROPA
at the J\"ulich Supercomputing Centre (FZ-J); we thank the center for their assistance.
Furthermore we acknowledge use of computing resources at the Danish
Center for Scientific Computing in Copenhagen, and at HLRS and FZ-J through
the DECI-6 grant SunFlare,
as well as funding from the European Commission's Seventh Framework Programme
(FP7/2007-2013) under the grant agreement SWIFF (project No. 263340, www.swiff.eu).

\end{document}